\documentclass[twocolumn,english,aps,prb,showpacs,superscriptaddress]{revtex4-1}
\usepackage[T1]{fontenc}
\usepackage[latin9]{inputenc}
\usepackage{color}
\usepackage{babel}
\usepackage{float}
\usepackage{amsmath}
\usepackage{amssymb}
\usepackage{graphicx}
\usepackage[colorlinks=true] {hyperref}
\hypersetup{linkcolor=blue,filecolor=blue,urlcolor=blue,citecolor=green}
\begin{document}
\title{Spread complexity and dynamical transition in multimode Bose-Einstein
condensates}
\author{Bozhen Zhou}
\affiliation{Beijing National Laboratory for Condensed Matter Physics, Institute
of Physics, Chinese Academy of Sciences, Beijing 100190, China}
\author{Shu Chen}
\email{Corresponding author: schen@iphy.ac.cn }

\affiliation{Beijing National Laboratory for Condensed Matter Physics, Institute
of Physics, Chinese Academy of Sciences, Beijing 100190, China}
\affiliation{School of Physical Sciences, University of Chinese Academy of Sciences,
Beijing 100049, China }
\begin{abstract}
We study the spread complexity in two-mode Bose-Einstein condensations
and unveil that the long-time average of the spread complexity $\overline{C}_{K}$
can probe the dynamical transition from self-trapping to Josephson
oscillation. When the parameter $\omega$ increases over a critical
value $\omega_{c}$, we reveal that the spread complexity exhibits
a sharp transition from lower to higher value, with the corresponding
phase space trajectory changing from self-trapping to Josephson oscillation.
Moreover, we scrutinize the eigen-spectrum and uncover the relation
between the dynamical transition and the excited state quantum phase
transition, which is characterized by the emergence of singularity
in the density of states at critical energy $E_{c}$. In the thermodynamical
limit, the cross point of $E_{c}(\omega)$ and the initial energy
$E_{0}(\omega)$ determines the dynamical transition point $\omega_{c}$.
Furthermore, we show that the different dynamical
behavior for the initial state at a fixed point can be distinguished
by the long-time average of the spread complexity, when the fixed
point changes from unstable to stable. Finally, we
also examine the sensitivity of $\overline{C}_{K}$ for the triple-well bosonic model which exibits
the transition from chaotic dynamics to regular dynamics.
\end{abstract}
\maketitle

\section{INTRODUCTION}

As a paradigmatic platform for investigating intriguing dynamical
phenomena, two-mode Bose-Einstein condensates (BECs) have attracted
intensive studies in past decades \citep{Andrews1997Science,Milburn1997PRA,Smerzi1997PRL,Leggett2001RMP,Micheli2003PRA,Zhou2003JPAM,Mahmud2005PRA,LiangCG,Tonel2005JPAM,Theocharis2006PRE}.
In a two-mode approximation, a two-component BEC or a BEC trapped
in a double-well potential can be effectively described by a two-mode
or two-site Bose-Hubbard model \citep{Milburn1997PRA,Smerzi1997PRL,Leggett2001RMP,Anglin2001PRA,Korsch2007PRA,Links2006AHP,Folling2007Nature,Julia2010PRA,Rubeni2017PRA,Boukobza2009PRL},
which is equivalently represented by a large spin model, known as the
Lipkin-Meshkov-Glick (LMG) model \citep{LMG1965,Ma2009PRE} in a different
parameter region. The two-mode BECs exhibit rich dynamical behaviors,
such as Josephson oscillation \citep{Milburn1997PRA,Smerzi1997PRL}
and self-trapping \citep{Anglin2001PRA,Smerzi1999,Albiez2005PRL,Liu2006PRA},
which have been studied in the scheme of the nonlinear Schr$\ddot{o}$dinger equation
and the Bose-Hubbard model. On the other hand, the LMG model is a
prototypical model for studying quantum phase transition and excited
state phase transition \citep{Corps2021PRL,Perez2011PRA,Wang2017PRE,Santos2016PRA,Gamito2022PRE,Dusuel2004PRL,Relano2009}.
It has been widely applied to study equilibrium and nonequilibrium
properties of quantum many-body systems \citep{Corps2022PRB,Corps2023PRL,Chinni2021PRR,LMGdyn}.

In past years, quenching a quantum system far from equilibrium was
used to unveil exotic dynamical phenomena, e.g., the long-time
average of order parameter changes nonanalytically at a dynamical
transition point \citep{Sciolla2013PRB,Piccitto2019RPB,Muniz2020Nature,DPT1,DPT2,DPT3},
and a series of non-analytical zero points at critical times are present
in the Loschmidt echo during time evolution \citep{Heyl2013PRL,Karrasch2013PRB,Jurcevic2017PRL,Heyl2019,Lang2018,Sirker,Canovi}.
Both non-analytical behaviours relate to the intrinsic property
of the system and belong to the class of dynamical phase transition.
Usually, fynamical properties of a many-body system
need to be diagnosed by various quantities from different perspectives to be fully understood.
The concept of complexity is such a quantity that has been used to
characterize the speed of the quantum evolution \citep{Brown2017PRD,Balasubramanian2020JHEP,Balasubramanian2021JHEP}.
In terms of complexity, the universal properties of operator growth
can be seen in the Lanczos coefficients after expanding the operator
in Krylov basis \citep{Parker2019PRX}. Furthermore, the properties
of a quantum phase are also rooted in the complexity of a state during
the dynamical evolution \citep{Sokolov2008PRE,Balasubramanian2022PRD,Caputa2022PRB,Afrasiar2023JSM,Caputa2023JHEP,Pal2023PRB,Pal2023,Zhai2023},
which can be obtained from the quantity named spread complexity. Motivated
by these progresses, it is interesting to explore whether complexity
can be used as an efficient probe to distinguish different dynamical
behaviors in many-body systems.

In this paper, we utilize the spread complexity $C_{K}$ in the Krylov
basis to characterize the dynamical transition occurring in two-mode
BECs. Usually, this dynamical transition is characterized by the non-analyticity
of the long-time average of the order parameters in quench dynamics.
Here we find that the long-time average of the spread complexity $\overline{C}_{K}$
can characterize the dynamical transition in two-mode BECs, consistent
with the result obtained from the analysis of dynamical order parameter.
It exhibits a transition from the lower complexity to the higher complexity
as the phase space trajectory changes from self-trapping to Josephson
oscillations. Although the semiclassical phase space dynamics \citep{Bohigas1993}
provides instructive understanding of the dependence of dynamical
transition on the choice of initial state, it is still elusive to
understand the role of the eigenspectrum of the underlying Hamiltonian
which governs the dynamical evolution. By examining the overlap between
the initial state and the eigenstates of the Hamiltonian, we demonstrate
that the dynamical behaviour of the quantum system is dominated by a
small portion of the eigenstates with energy near the initial state
energy. To deepen our understanding, we analyze the structure of the spectrum
and uncover the relation of dynamical transition to the excited
state quantum phase transition \citep{EQPT,EQPT2}, which is characterized
by the emergence of singularity in the density of states at critical
energy $E_{c}$ \citep{EQPT,EQPT2,Santos2016PRA,Relano2009,Perez2011PRA}.
Under semiclassical approximation, the critical energy corresponds
to the energy of a saddle point, which separates the degenerate region
and non-degenerate region. When the parameter $\omega$ increases
over a threshold $\omega_{th}$, the saddle point becomes a maximum,
and the corresponding dynamics changes dramatically. By studying the
dynamics with the initial state at this fixed point, we show that
the spread complexity $C_{K}(t)$ exhibits quite different behavior
in the region above or below $\omega_{th}$, and the transition can
be characterized by the long-time average of the spread complexity.
Finally, we also consider the triple-well bosonic model, in which the chaotic dynamics and regular dynamics have been well studied\citep{Hiller2009PRA,Rautenberg2020PRA,Santos2022PRE,Castro2024PRA}. 
We find that $\overline{C}_{K}$ can also characterize chaotic-regular transition in such a system. 

The rest of the paper is organized as follows. In Sec. II, we briefly
introduce the spread complexity and derive the expression of long-time
average of the spread complexity. In Sec. III A, we study the dynamical
transition in two-mode BECs and demonstrate that the different dynamical
behaviors in the self-trapping regime and Josephson oscillation regime
can be characterized by the sharp change of the long-time average
of spreading complexity. In Sec. III B, we study the dynamical behavior of spreading
complexity around a fixed point and demonstrate that different behavior
in the region above or below $\omega_{th}$ can be characterized by
the long-time average of the spread complexity. In Sec. III C, we unveil the relation of
dynamical transition with the spectrum structure of the underlying
Hamiltonian. In Sec. IV, we study the triple-well bosonic model and show that the
long-time average of spreading complexity can be used to distinguish the chaotic
dynamics and regular dynamics. A summary is given in Sec. V.

\section{Long-time average of the spread complexity}

Consider a quantum system with a time-independent Hamiltonian $H$.
For convenience, we set $\hbar=1$. Then the time evolution of a state
$|\psi(t)\rangle$ is governed by $|\psi(t)\rangle=e^{-iHt}|\psi(0)\rangle$.
Expanding the right hand side in power series, we get
\begin{equation}
|\psi(t)\rangle=\sum_{k=0}^{\infty}\frac{(-it)^{k}}{k!}|\psi_{k}\rangle,
\end{equation}
where $|\psi_{k}\rangle=H^{k}|\psi(0)\rangle$. Then applying the
Gram--Schmidt process to the set of vectors $\left\{ |\psi_{0}\rangle,|\psi_{1}\rangle...|\psi_{k}\rangle\right\} $,
it generates an orthogonal basis $\mathcal{K}\doteq\left\{ |K_{0}\rangle,|K_{1}\rangle...|K_{k}\rangle\right\} $
with $|K_{0}\rangle\equiv|\psi_{0}\rangle$. The basis $\mathcal{K}$
is called the Krylov basis \citep{Balasubramanian2022PRD,Afrasiar2023JSM}.
In this paper, we consider the complete orthonormal basis of the Hilbert
space with the maximal value of $k$ being $\mathcal{D}-1$, where
$\mathcal{D}$ is the dimension of the Hamiltonian $H$. The full
algorithm is described as following: After choosing the initial state
$|K_{0}\rangle$, the subsequent Krylov bases can be obtained recursively
by the following algorithm:
\begin{align}
|\tilde{\psi}_{n}\rangle & =|\psi_{n}\rangle-\sum_{k=0}^{n-1}\langle K_{k}|\psi_{n}\rangle|K_{k}\rangle,\nonumber \\
b_{n} & =\sqrt{\langle\tilde{\psi}_{n}|\tilde{\psi}_{n}\rangle},\\
|K_{n}\rangle & =\frac{1}{b_{n}}|\tilde{\psi}_{n}\rangle.\nonumber
\end{align}
Then the Hamiltonian becomes a tridiagonal form in the Krylov basis
$\mathcal{K}$:
\begin{equation}
H_{\mathcal{K}}=\begin{bmatrix}a_{0} & b_{1} & 0 & 0\\
b_{1} & a_{1} & \ddots & 0\\
0 & \ddots & \ddots & b_{k}\\
0 & 0 & b_{k} & a_{k}
\end{bmatrix},
\end{equation}
where $a_{k}\equiv\langle K_{k}|H|K_{k}\rangle$ and $b_{k}$ are
also called Lanczos coefficients \citep{Viswanath1950}. In our numerical
calculation, we use the MPLAPACK \citep{Nakata2021arXiv} library
to perform the arbitrary precision computation.

Using the Krylov basis, we can define the spread complexity as \citep{Balasubramanian2022PRD}
\begin{equation}
C_{K}(t)=\sum_{k=0}^{\mathcal{D}-1}k\left|\langle K_{k}|\psi(t)\rangle\right|^{2}.
\end{equation}
The spread complexity quantifies the degree of complex of the initial
state $|\psi(0)\rangle$ during the time evolution. It can be observed
that the return probability is defined as $\mathcal{L}(t)=\left|\langle K_{0}|\psi(t)\rangle\right|^{2}$.
The return probability is also known as Loschmidt echo and has been
widely studied in the non-equilibrium system \citep{Heyl2013PRL,Karrasch2013PRB,Heyl2019,Zhou2019PRB,ZengYM,ZhouBZ}.
In this paper, we focus on the long-time average of the spread complexity
\begin{equation}
\overline{C}_{K}\equiv\lim_{T\rightarrow\infty}\frac{1}{T}\int_{0}^{T}C_{K}(t)dt.
\end{equation}
Inserting the complete set of energy eigenstates, we get
\begin{align}
\overline{C}_{K} & =\sum_{k=0}^{\mathcal{D}-1}k\sum_{n=1}^{\mathcal{D}}\left|\alpha_{kn}\right|^{2}\left|\alpha_{0n}\right|^{2},\label{eq:=000020Ckb_ana}
\end{align}
where the coefficients are given by $\alpha_{kn}=\langle K_{k}|\phi_{n}\rangle$
with $H|\phi_{n}\rangle=E_{n}|\phi_{n}\rangle$.

\section{Two-mode Bose-Einstein condensates and its spread complexity}

Now, we consider a two-mode Bose-Einstein condensates with the Hamiltonian
described by \citep{Leggett2001RMP,Micheli2003PRA,Tonel2005JPAMG,Zibold2010PRL,Fan2012PRA}:
\begin{equation}
H=\frac{2\chi}{N}\hat{S}_{z}^{2}+\omega\hat{S}_{x},\label{eq:H_LMG}
\end{equation}
where $\chi$ is atom-atom interaction and $\omega$ is the Rabi frequency
of the external field interacting with the condensate. For the sake
of convenience, we set $\chi=1$ as the unit of energy. The angular-momentum
operators $\hat{S}_{x}$, $\hat{S}_{y}$ and $\hat{S}_{z}$ are the
Schwinger pseudospin operators:
\begin{equation}
\begin{cases}
\hat{S}_{x}=\frac{1}{2}\left(\hat{a}_{1}^{\dagger}\hat{a}_{2}+\hat{a}_{2}^{\dagger}\hat{a}_{1}\right)\\
\hat{S}_{y}=\frac{i}{2}\left(\hat{a}_{2}^{\dagger}\hat{a}_{1}-\hat{a}_{1}^{\dagger}\hat{a}_{2}\right)\\
\hat{S}_{z}=\frac{1}{2}\left(\hat{a}_{1}^{\dagger}\hat{a}_{1}-\hat{a}_{2}^{\dagger}\hat{a}_{2}\right)
\end{cases}
\end{equation}
where $\hat{a}^{\dagger}$ and $\hat{a}$ is bosonic creation and
annihilation operator, respectively. This many-particle Hamiltonian
is closely related to the original LMG model \citep{LMG1965}, for
which however the parameter $\chi$ is negative.

\subsection{Long time average of the spread complexity and dynamical transition}

\begin{figure}[h]
\begin{centering}
\includegraphics[scale=0.12]{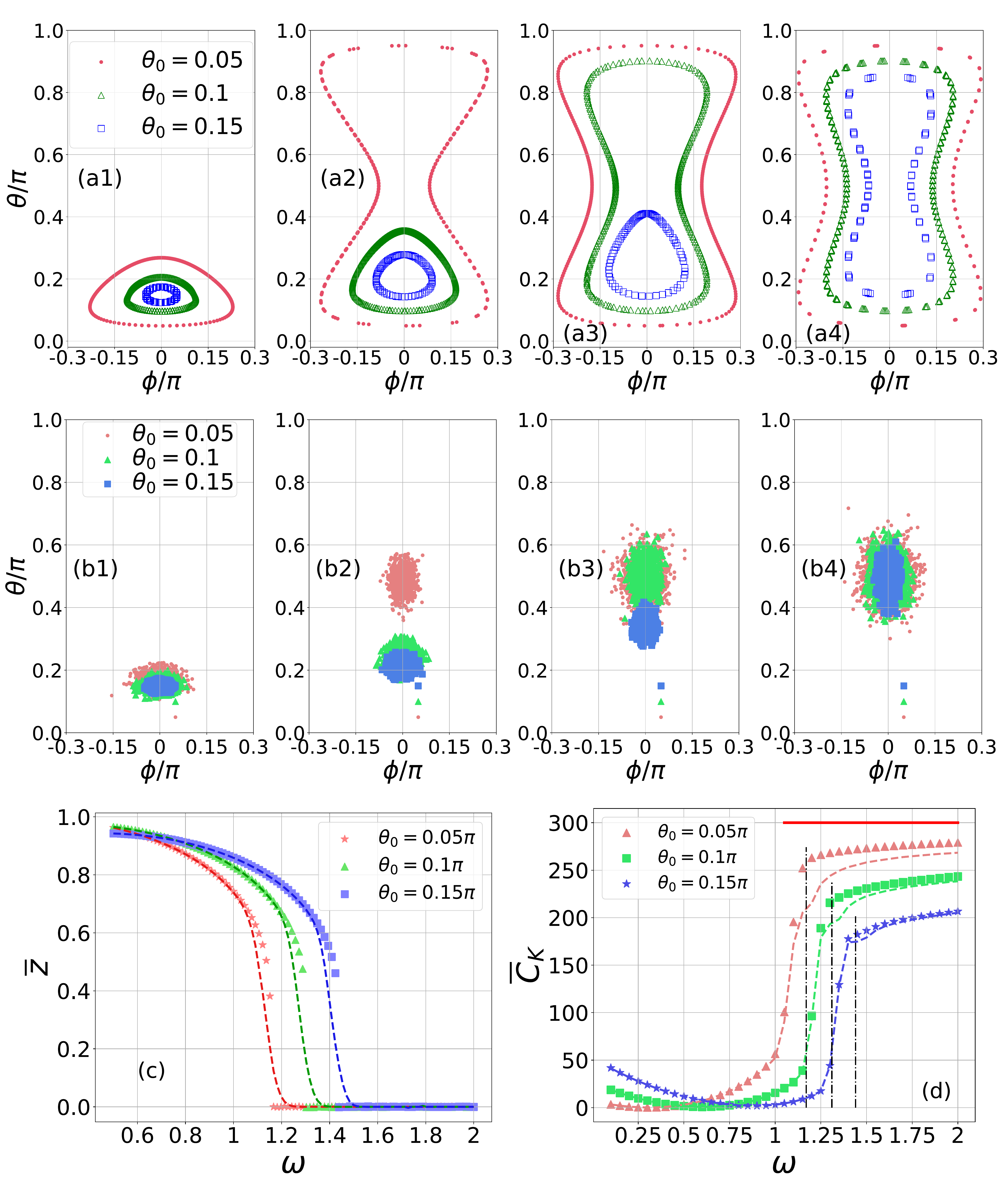}
\par\end{centering}
\caption{(a) Classical trajectories of the semi-classical model and (b) quantum
trajectories for different initial states in the $\theta-\phi$ plane.
The parameter are (a1)(b1) $\omega=0.9$; (a2)(b2) $\omega=1.2$; (a3)(b3)
$\omega=1.4$; (a4)(b4) $\omega=1.5$. (c) $\overline{z}$ vs
$\omega$. The corresponding dashed lines are obtained via the calculation
of $2\bar{S}_{z}/N$ for $N=600$. (d) $\overline{C}_{K}$ vs
$\omega$ with $\phi_{0}=0.05\pi$ for $N=600$. The dashed lines
in (d) are summing over within the energy window $\epsilon\in[E_{0}-2\delta E,E_{0}+2\delta E]$.
Red-bolded line corresponds to the maximally delocalized state for
$N=600$. Three dash-dotted lines are corresponding to three transition
points in (c). \protect\label{fig:traj_LMG}}
\end{figure}

Under the semi-classical approximation with $N\gg1$, angular-momentum
operators $\vec{S}$ can be replaced by $\vec{S}\rightarrow\frac{N}{2}\left(\sin\theta\cos\phi,\sin\theta\sin\phi,\cos\theta\right).$
Then we can obtain the equations of motion via the Heisenberg equations
of motion:
\begin{align}
\dot{\theta} & =-\omega\sin\phi,\\
\dot{\phi} & =2\chi\cos\theta-\omega\cot\theta\cos\phi.
\end{align}
The classical dynamics has been studied in the previous works which
showed the dynamical transition between self-trapped trajectory and
Josephson oscillation trajectory. Here, we demonstrate the classical
trajectories in Figs. \ref{fig:traj_LMG}(a1)$\sim$(a4), in which
we consider three initial values with $\phi_{0}=0.05\pi$ and $\theta_{0}=0.05\pi,0.1\pi,0.15\pi$.
The four figures corresponding to four different $\omega$ and their
trajectories form closed orbits. It can be found that three trajectories
show the self-trapped behaviour for small $\omega$. As $\omega$
increases, the trajectories for different initial state sequentially
become Josephson oscillation. Such as in Fig. \ref{fig:traj_LMG}(a2)
for $\omega=1.2$, only the trajectory with $\left(\theta_{0}=0.05\pi,\phi_{0}=0.05\pi\right)$
transition to the Josephson oscillation and others remain self-trapped
behaviour. However, in Fig. \ref{fig:traj_LMG}(a3) for $\omega=1.4$,
only the trajectory with $\left(\theta_{0}=0.15\pi,\phi_{0}=0.05\pi\right)$
remains in the self-trapped regime. The dynamical transition of the
classical trajectory can be captured by the order parameter $\bar{z}=\frac{1}{t}\int_{0}^{t}z(\tau)d\tau$
which is the time average of the canonical coordinate $z\equiv\cos\theta$.
In Fig. \ref{fig:traj_LMG}(c), we demonstrate the value of $\bar{z}$
with respect to $\omega$ for three different initial values. Here
we carry out the time average from 0 to 1000. It can be found that
$\bar{z}$ has a non-zero value for self-trapped trajectories but
approaches zero for Josephson oscillation trajectories.

For quantum dynamics, we choose the coherent spin state (CSS) as
the initial state \citep{Micheli2003PRA,Zhang2021PRB}. This state is
given by
\begin{equation}
|\theta_{0},\phi_{0}\rangle=e^{-iS_{z}\phi_{0}}e^{-iS_{y}\theta_{0}}|\frac{N}{2},\frac{N}{2}\rangle,
\end{equation}
where $|\frac{N}{2},\frac{N}{2}\rangle$ is the highest-weight state
of the SU(2) group with spin $\frac{N}{2}$ and $\langle S_{z}\rangle=\frac{N}{2}$.
The CSS takes its maximum polarization in the direction $\left(\theta_{0},\phi_{0}\right)$.
Such a choice of the initial state is relevant to analyze the classical-quantum
correspondence. For quantum trajectory, we calculate the time evolved
state $|\psi(t)\rangle=e^{-iHt}|\psi_{0}\rangle$ with $|\psi_{0}\rangle\equiv|\theta_{0},\phi_{0}\rangle$
and corresponding time dependent expectation value $\langle S_{x}(t)\rangle$,
$\langle S_{y}(t)\rangle$ and $\langle S_{z}(t)\rangle$. Then we
transform $\left(\langle S_{x}(t)\rangle,\langle S_{y}(t)\rangle,\langle S_{z}(t)\rangle\right)$
into sphere coordinate $\left(R\sin\theta\cos\phi,R\sin\theta\sin\phi,R\cos\theta\right)$
where $R^{2}=\langle S_{x}(t)\rangle^{2}+\langle S_{y}(t)\rangle^{2}+\langle S_{z}(t)\rangle^{2}$.
Similar to the classical trajectory, we present the quantum trajectories
in the $\theta-\phi$ plane, as shown in Figs. \ref{fig:traj_LMG}(b1)$\sim$(b4).
The parameters are the same as in Figs. \ref{fig:traj_LMG}(a1)$\sim$(a4).
It can be observed that the areas of the quantum trajectories are
related to the classical trajectories. Particularly, the initial state
dependent dynamics transition can also be observed in the quantum
trajectory. Similar to the order parameter $\bar{z}$, we can choose
the order parameter $\bar{S}_{z}=\underset{t\rightarrow\infty}{\lim}\frac{1}{t}\int_{0}^{t}\langle\psi_{0}|S_{z}(t)|\psi_{0}\rangle d\tau$
in quantum dynamics. In Fig. \ref{fig:traj_LMG}(c), we show the values
of $\bar{z}\equiv\frac{2}{N}\bar{S}_{z}$ by dashed lines and they
are similar to the semi-classical ones except that the transition
points are smoothed by the finite-size effect.
\begin{figure}[H]
\begin{centering}
\includegraphics[scale=0.14]{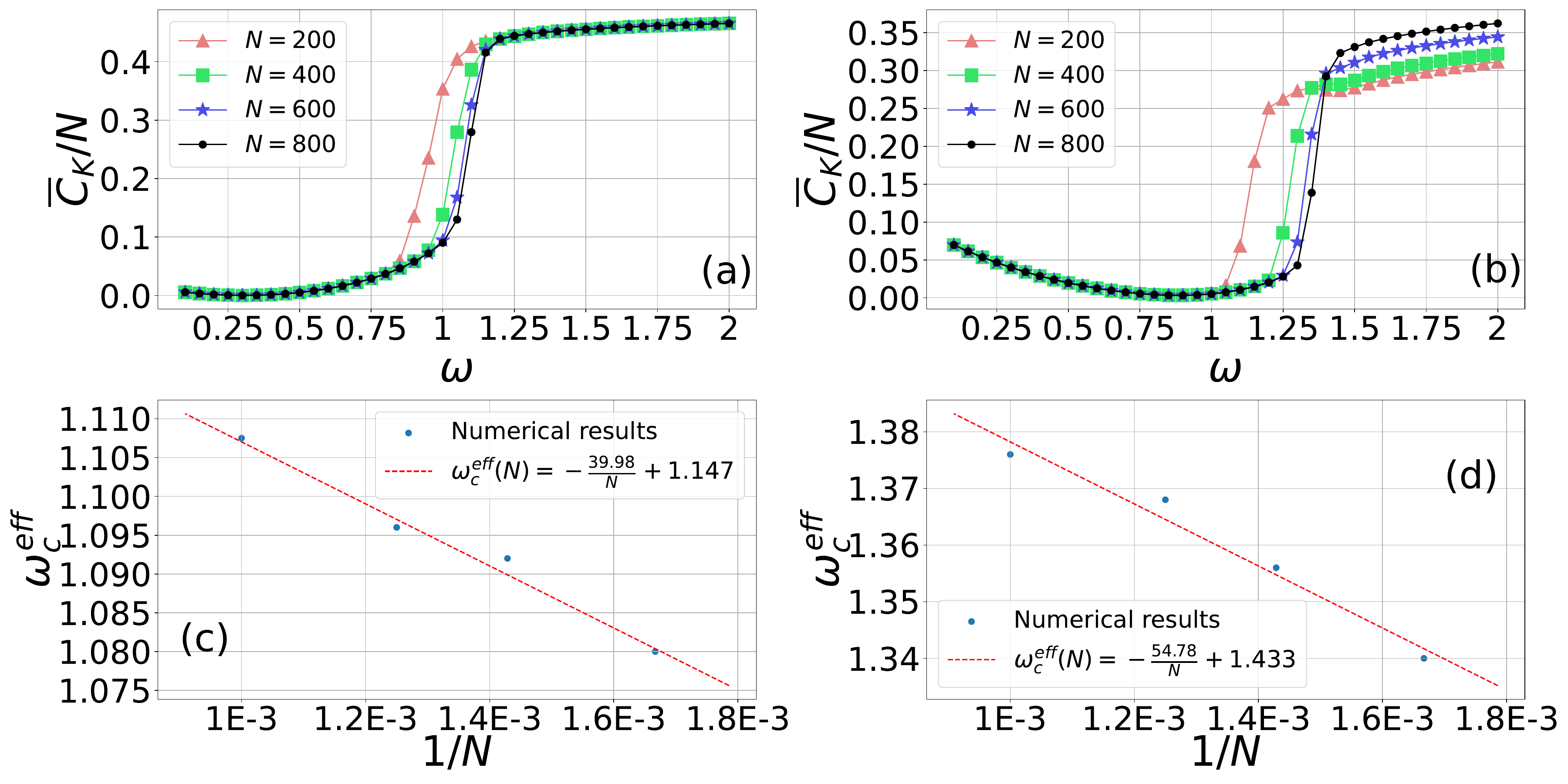}
\par\end{centering}
\caption{(a)(b) $\overline{C}_{K}$ vs $\omega$ for different system sizes.
(c)(d) Finite-size scaling for the effective transition point $\omega_{c}^{\text{eff}}$.
The initial states are (a)(c) $\theta_{0}=0.05\pi,\phi_{0}=0.05\pi$;
(b)(d) $\theta_{0}=0.15\pi,\phi_{0}=0.05\pi$ . \protect\label{fig:FSA_w}}
\end{figure}

The different distribution of the quantum trajectories can be characterized
by the long-time average of the spread complexity $\overline{C}_{K}$
which is displayed in Fig. \ref{fig:traj_LMG}(d). It can be found
that the larger accessible area of trajectory corresponds to the larger
value of the spread complexity and vice versa. The connection between
quantum trajectories and the spread complexity is intuitive, 
as different Krylov bases distribute in different regions of the phase
space. More evidence can be seen in Appendix A where we show the Husimi
function in the phase space for the $k$-th Krylov basis. For the
self-trapped trajectories, the time evolved state is constrained in
a small area of the phase space. In the perspective of Krylov space,
the self-trapped trajectory is dynamically localized near the space
of the initial Krylov state $|K_{0}\rangle$. For the extremely localized
case, the dynamics is frozen at initial state and we have $\left|\langle K_{k}|\psi(t)\rangle\right|^{2}\approx\delta_{k0}$
and $C_{K}(t)\approx0$. However, for the Josephson oscillation trajectories,
the time evolved state extends in the phase space and widely distributes
in the Krylov space. Considering the maximally delocalized state $|\psi_{\text{d}}\rangle$
in Krylov space, we have $\left|\langle K_{k}|\psi_{\text{d}}\rangle\right|^{2}\rightarrow\frac{1}{\mathcal{D}}$
and $\overline{C}_{K}\approx\frac{\mathcal{D}-1}{2}=\frac{N}{2}$.
The value $\frac{N}{2}$ is drawn by the red-bolded line for $N=600$
in Fig. \ref{fig:traj_LMG}(d). It can be seen that $\overline{C}_{K}$
is closer to the value $\frac{N}{2}$ for the larger area of quantum
trajectory.

To get an intuitive insight how the wavef unction spreads out, we also consider the inverse
participation ratio of the time evolved state which is defined
as
\begin{equation}
\mathcal{I}_{K}(t)=\sum_{k}\left|\langle K_{k}|\psi(t)\rangle\right|^{4}.
\end{equation}
We calculate the  inverse participation ratio and return probability of the time evolved state
and show the results in the Appendix B. It is shown that the inverse participation ratio exhibits quite different behaviors in the self-trapped and Josephson oscillation regions.

Next we carry out the finite-size analysis on the transition point.
To determine the transition point, we differentiate the function $\overline{C}_{K}$
with respect to $\omega$ and label the location of the maximum of
$\frac{\partial\overline{C}_{K}}{\partial\omega}$ as $\omega_{c}^{\text{eff}}$,
which is size dependent. We show $\omega_{c}^{\text{eff}}$ for different
size in Fig. \ref{fig:FSA_w}(c) and Fig. \ref{fig:FSA_w}(d). Further
linear fitting the results of $\omega_{c}^{\text{eff}}$ indicate
the transition point at large $N$ limit is $\omega_{c}^{\text{eff}}(N\rightarrow\infty)\approx1.147$
for $\theta_{0}=0.05\pi$ and $\omega_{c}^{\text{eff}}(N\rightarrow\infty)\approx1.433$
for $\theta_{0}=0.15\pi$. These converged values are close to the
transition points $\omega_{c}\approx1.167$ and $1.439$
present in the order parameter $\bar{z}$ in Fig. \ref{fig:traj_LMG}(c).

\subsection{Dynamical behaviour of $C_{K}(t)$ around a fixed point}

\begin{figure}[H]
\begin{centering}
\includegraphics[scale=0.17]{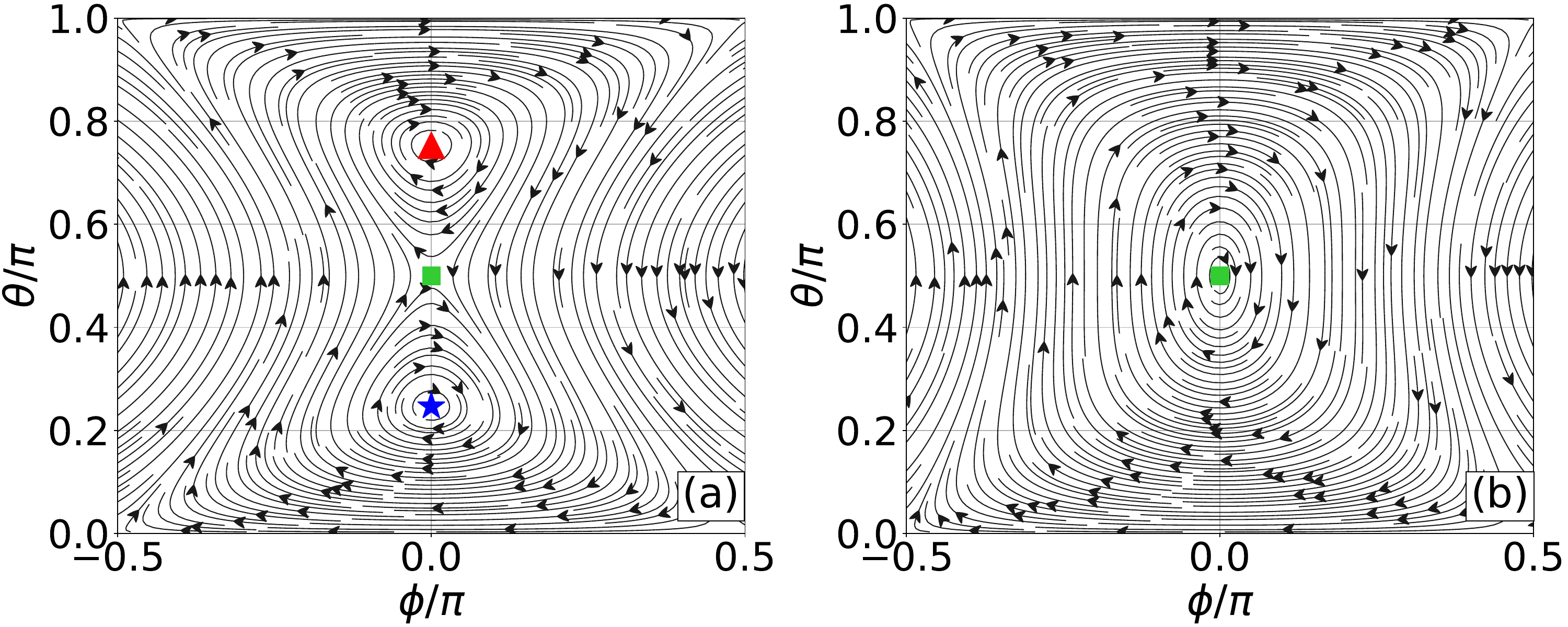}
\par\end{centering}
\caption{Tangent vector field of the equation of motion of the semi-classical model
for (a) $\omega=1.4$ and (b) $\omega=2.5$. Three symbols denote
the three fixed points: star symbol $\left(\theta=\arcsin\frac{\omega}{2},\phi=0\right)$,
triangular symbol $\left(\theta=\pi-\arcsin\frac{\omega}{2},\phi=0\right)$
and square symbol $\left(\theta=\frac{\pi}{2},\phi=0\right)$.\protect\label{fig:fixed_ps}}
\end{figure}

Now we study the dynamics around the fixed point $\left(\theta=\frac{\pi}{2},\phi=0\right)$.
In the regime of $\omega<2$, $\left(\theta=\frac{\pi}{2},\phi=0\right)$
is a saddle point from the perspective of energy surface, denoted
by the square symbol in Fig. \ref{fig:fixed_ps} (a). Besides, there
are two degenerate maxima: $\left(\theta=\arcsin\frac{\omega}{2},\phi=0\right)$
and $\left(\theta=\pi-\arcsin\frac{\omega}{2},\phi=0\right)$, denoted
by the star symbol and the triangular symbol in Fig. \ref{fig:fixed_ps}
(a) for $\omega=1.4$, respectively. These two maxima merge into
one point $\left(\theta=\frac{\pi}{2},\phi=0\right)$ at $\omega=2$.
For $\omega>2$, there is only a maxima at $\left(\theta=\frac{\pi}{2},\phi=0\right)$,
as demonstrated in Fig. \ref{fig:fixed_ps} (b) for $\omega=2.5$.
When $\omega$ increases over the threshold $\omega_{th}=2$, the
trajectory is also dramatically changed, and the corresponding dynamics
changes from the Josephson oscillation to the Rabi oscillation \citep{Leggett2001RMP,Tonel2005JPAM}.
This transition can be characterized by the fixed point $\left(\theta=\frac{\pi}{2},\phi=0\right)$
whose Jacobian matrix is
\begin{equation}
\mathcal{J}=\begin{bmatrix}0 & 2-\omega\\
\omega & 0
\end{bmatrix}.
\end{equation}
The two eigenvalues of the matrix $\mathcal{J}$ are $\pm\sqrt{\omega(2-\omega)}$.
For $\omega\in(0,2)$, two eigenvalues are real number and mutually
opposite. So this fixed point is the unstable saddle point. As shown
in Fig. \ref{fig:fixed_ps}(a) for $\omega=1.4$, the tangent vector
is away from the fixed point. For $\omega\in[2,+\infty)$, two eigenvalues
are imaginary numbers. So the fixed point is stable and called the center\citep{Strogatz2014}
whose nearby trajectories are neither attracted to nor repelled from
the fixed point, as illustrated in Fig. \ref{fig:fixed_ps}(b). The
threshold point $\omega_{th}=2$ splits two qualitatively different
dynamical behavior, i.e., Josephson-type versus Rabi-type oscillation.

\begin{figure}[H]
\begin{centering}
\includegraphics[scale=0.12]{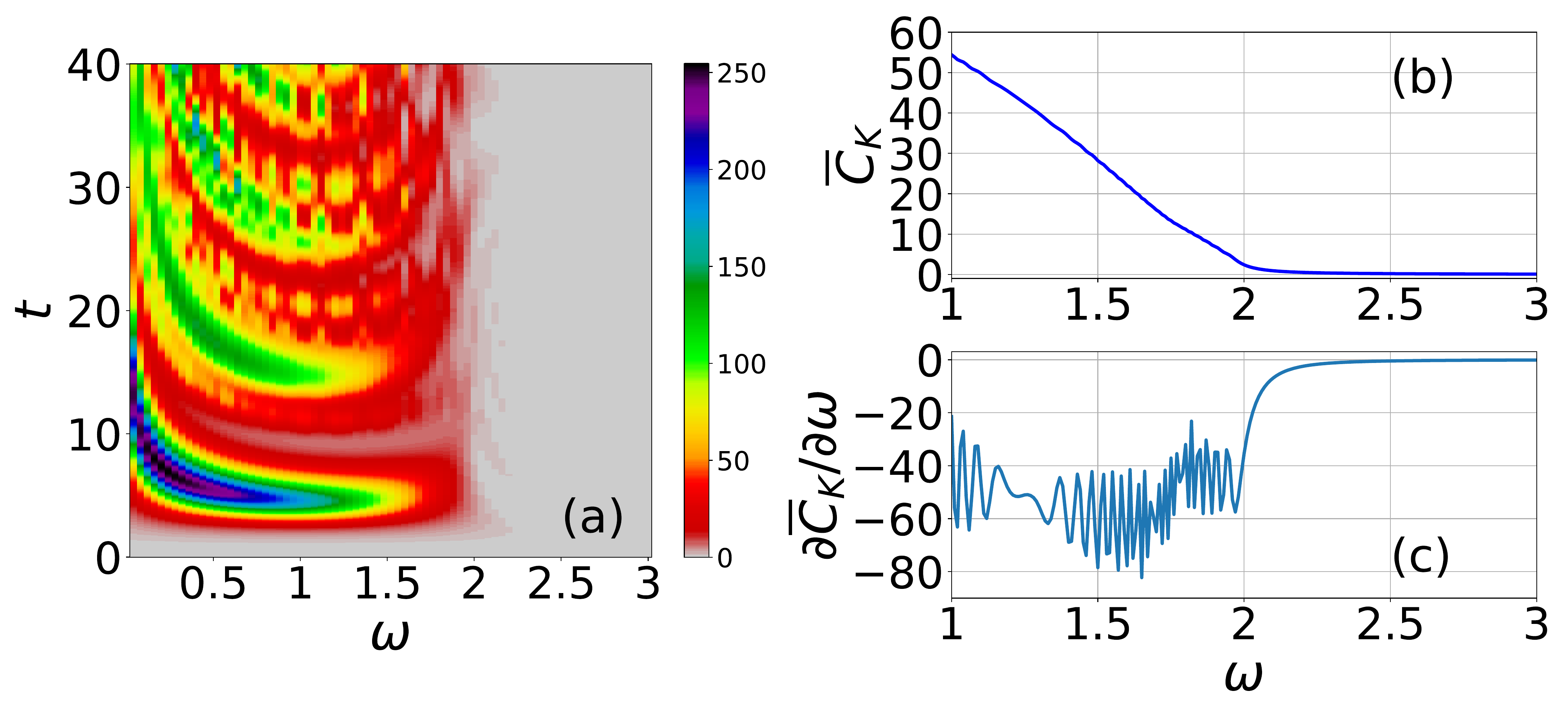}
\par\end{centering}
\caption{(a) Time evolution of $C_{K}(t)$ starting from the initial state
$|\theta_{0}=\frac{\pi}{2},\phi_{0}=0\rangle$ with respect to $\omega$.
(b) $\overline{C}_{K}$ with respect to $\omega$. (c) Derivative
of $\overline{C}_{K}$ with respect to $\omega$. The system size is $N=800$.
\protect\label{fig:saddle}}
\end{figure}

For the quantum system, it has been revealed the existence of exotic
dynamical behavior around unstable fixed points \citep{Cao2020PRL,Hummel2019PRL,Richter2023,Kidd2021,Bhattacharjee2022JHEP},
which is referred to as the scrambling characterized by the exponential
growth of the out-of-time order correlators. Setting the initial state
as $|\theta_{0}=\frac{\pi}{2},\phi_{0}=0\rangle$, we study how the
spread complexity changes with $\omega$. Here, we show the value
of the spread complexity $C_{K}(t)$ and its long-time average value
$\overline{C}_{K}$ in Fig. \ref{fig:saddle}(a) and Fig. \ref{fig:saddle}(b),
respectively. The dynamics of the $C_{K}(t)$ suggests that the initial
state would evolve to states far away from $|K_{0}\rangle$ for $\omega<2$,
but stays near the initial state for $\omega>2$. The dramatically
distinct behaviour presented in the $C_{K}(t)$ is also evidenced
by the long-time average $\overline{C}_{K}$. When $\omega>2$, $\overline{C}_{K}$
approaches to zero, as shown in Fig. \ref{fig:saddle}(b).

For the limit case with $\omega\rightarrow\infty$, the Hamiltonian
can be simplified as $H_{\omega\rightarrow\infty}=S_{x}$, and the
initial state $|\theta_{0}=\frac{\pi}{2},\phi_{0}=0\rangle$ is the
eigenstate of $H_{\omega\rightarrow\infty}$. After dropping a global
phase, $|\psi(t)\rangle\propto|\theta_{0}=\frac{\pi}{2},\phi_{0}=0\rangle$
is time independent in the large $\omega$ limit and the spread complexity
remains zero during the time evolution. For $\omega<2$, the initial
energy $E_{0}$($\omega)$ for the initial state $|\theta_{0}=\frac{\pi}{2},\phi_{0}=0\rangle$
equals to the critical energy $E_{c}(\omega)$ which separates the
self-trapped trajectory and Josephson-type trajectory.  The discussion of critical energy can be seen in Sec. III C. Additionally, the derivative of the $\overline{C}_{K}$
with respect to $\omega$ displays oscillation for $\omega<2$. The
oscillation originates from the quantum fluctuation near the critical
energy as the density of state exhibits local divergence.

\begin{figure}[H]
\begin{centering}
\includegraphics[scale=0.12]{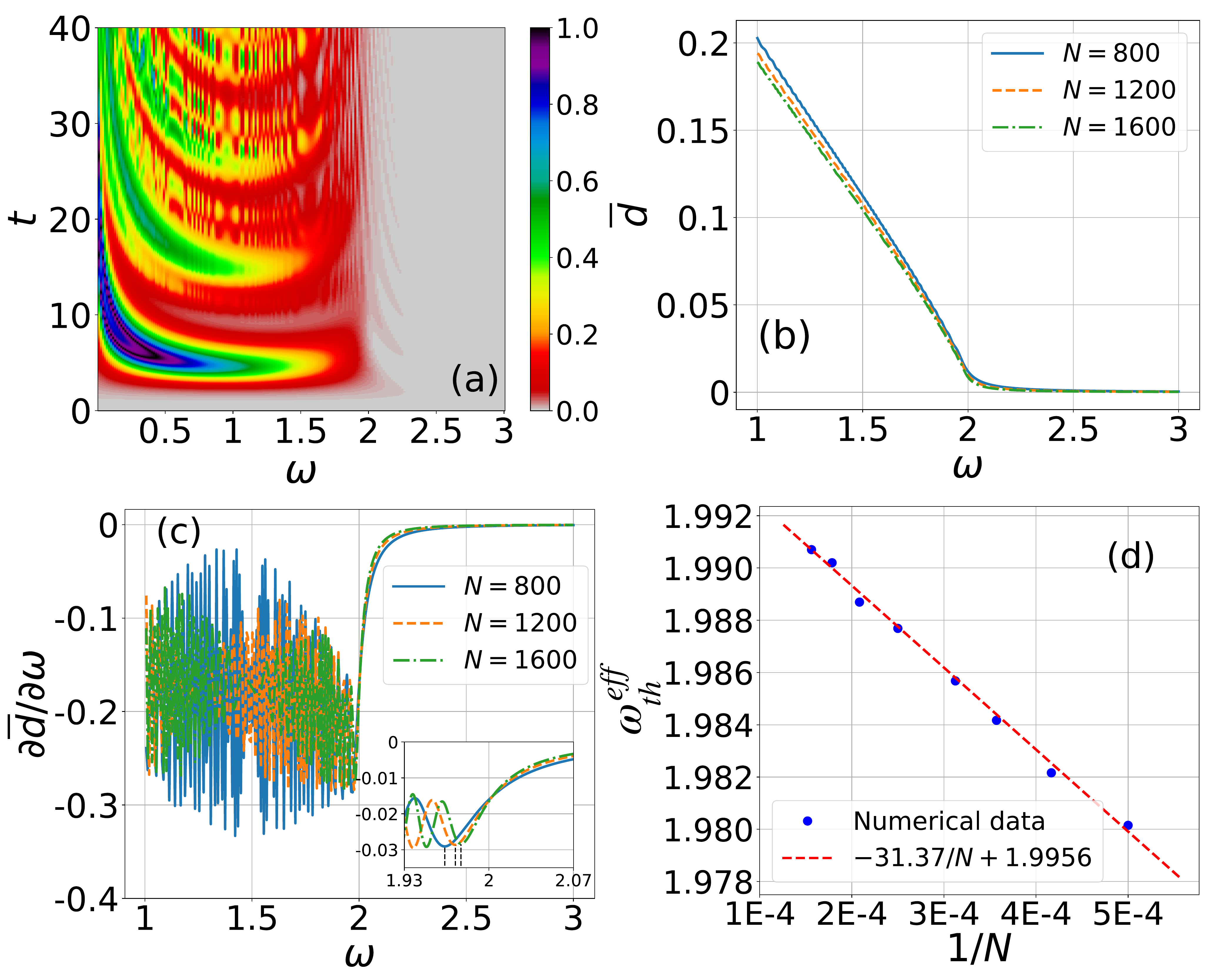}
\par\end{centering}
\caption{(a) Time evolution of $d(t)$ starting from the initial state $|\theta_{0}=\frac{\pi}{2},\phi_{0}=0\rangle$
with respect to $\omega$. (b) Long-time averaged values $\overline{d}$
with respect to $\omega$. (c) Derivative of $\overline{d}$ with
respect to $\omega$. The black dashed lines guide the minimal values
near $\omega=2$ which are used in finite-size scaling. (d) Finite
size scaling for the transition point. The system size is $N=800$
for (a)(b)(c). \protect\label{fig:saddle_RC}}
\end{figure}

For the trajectory starting from the fixed point $|\theta_{0}=\frac{\pi}{2},\phi_{0}=0\rangle$,
the dynamics of a classical state is frozen on the $\theta-\phi$
plane. However, the remaining radial coordinate $R$ of a quantum
trajectory is not conserved during the time evolution. Then, we can
define the distance of the time evolved state away from the initial
state in the phase space $R$ as
\begin{equation}
d(t)=\frac{2}{N}\left|R(t)-R(0)\right|.
\end{equation}
It can be expected that the distance $d(t)$ connects to the state
complexity $C_{K}(t)$ because they both measure the distance between the
time evolved state and the initial state. To see it clearly, we display
the short-time dynamics of $d(t)$ with respect to the parameter $\omega$
in the Fig. \ref{fig:saddle_RC}(a) and its long-time average value
$\overline{d}$ in Fig. \ref{fig:saddle_RC}(b). Comparing Figs. \ref{fig:saddle}(a)(b)
and Fig. \ref{fig:saddle_RC}(a)(b), it can be seen that the dynamical
behaviour of $d(t)$ is very similar to the dynamical behaviour of
$C_{K}(t)$. The derivative of the $\overline{d}$ with respect to
$\omega$ also shows oscillation for $\omega<2$. The similarity between
$d(t)$ and $C_{K}(t)$ results from the fact that both of them quantify the
distance between the time evolved state and the initial state. Here, we
can label the location of the minimum of $\frac{\partial\overline{d}}{\partial\omega}$
near $\omega=2$ as an effective transition point $\omega_{th}^{\text{eff }}$,
which is guided by the black dashed lines in the insert of Fig. \ref{fig:saddle_RC}(c).
It can be seen that $\omega_{th}^{\text{eff}}$ separates the oscillation
and non-oscillation regime of $\frac{\partial\overline{d}}{\partial\omega}$.
From the result of finite-size scaling shown in Fig. \ref{fig:saddle_RC}(d),
we can obtain $\omega_{th}^{\text{eff }}(N\rightarrow\infty)\approx1.996$,
which is approximately equal to the threshold point $\omega_{th}=2$.
The transition point in $\overline{C}_{K}$ is the same as $\overline{d}$
because they share the same physical origin.

\subsection{Relation to the spectrum structure}

\begin{figure*}
\centering \includegraphics[scale=0.22]{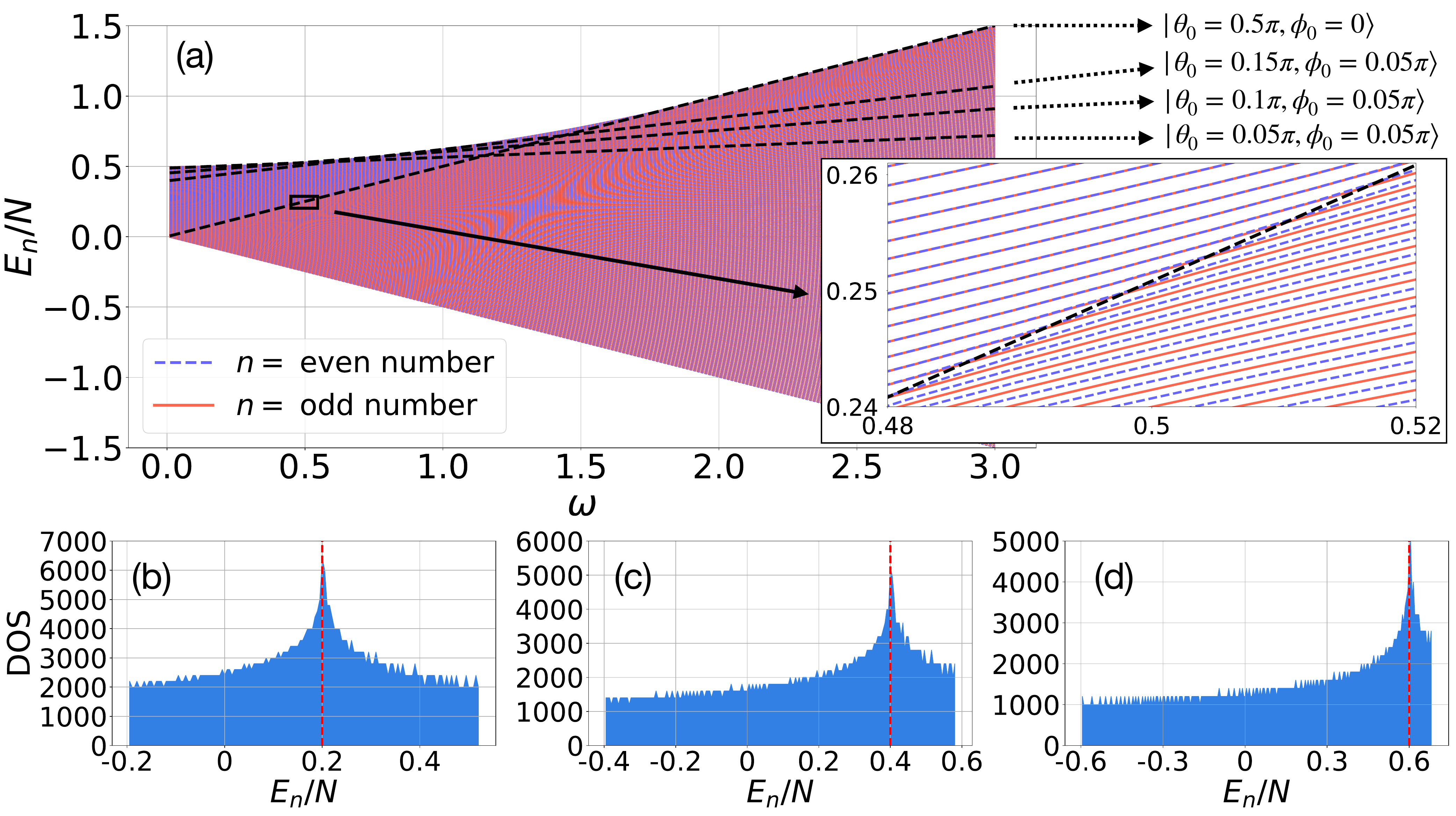}
\caption{(a) The energy spectrum with respect to $\omega$. The dashed black
lines are the initial energy $E_{0}$ corresponding to different initial
states. For clarity we have used a small system $N=600$. In the region
of $\omega<\omega_{th}$ ($\omega_{th}=2$), there exists excited
state quantum phase transition. In this region, the eigenstates are
separated by the critical energy $E_{c}$, at which the density of
states is divergent in the thermodynamical limit. The inset of (a)
demonstrates that states above $E_{c}$ are doubly degenerate, whereas
states below $E_{c}$ are non-degenerate. Densities of states are shown for (b)
$\omega=0.4$, (c) $\omega=0.8$ and (d) $\omega=1.2$ with $N=2000$.
The red dashed lines guide the value of critical energy $E_{c}\approx\frac{N}{2}\omega$
for $N\rightarrow\infty$.\protect\label{fig:E}}
\end{figure*}

To unveil the relation of dynamical transition to the
spectrum structure, we examine the eigen-spectrum of two-mode BECs with respect to $\omega$.
Here, we sort the eigenvalues in such a way that $E_{1}\leq E_{2}\leq\cdots\leq E_{\mathcal{D}}$
and divide the set $\{E_{n}\}$ into two subsets $\left\{ E_{n\in\text{even}}\right\} $
and $\left\{ E_{n\in\text{odd}}\right\} $. In Fig. (\ref{fig:E})(a),
we display the values of $\left\{ E_{n\in\text{even}}\right\} $ and
$\left\{ E_{n\in\text{odd}}\right\} $, corresponding to the blue
dashed lines and red solid lines, respectively. Further considering
the initial energy $E_{0}(\omega)\equiv\langle\psi_{0}|H(\omega)|\psi_{0}\rangle$,
it can be found that the initial energies $E_{0}(\omega)$ corresponding
to three initial states discussed previously go from the doubly degenerate
regime to the non-degenerate regime as $\omega$ increases. The critical
energy $E_{c}(\omega)$ separates the doubly degenerate regime from
the non-degenerate regime in the thermodynamic limit. As depicted
in Figs. \ref{fig:E}(b)(c)(d), the critical energy $E_{c}(\omega)$
can be evidenced by the local divergence in the density of states
\citep{Santos2016PRA}. While states below $E_{c}$ are non-degenerate,
the states above $E_{c}$ are degenerate in the thermodynamic limit.
For a finite-size system, it should be noted that the gap of the doubly
degenerate energy is exponentially small. With the increasing of $\omega$,
the region of doubly degenerate energy shrinks and eventually vanishes at
$\omega_{th}=2$. The critical energy $E_{c}(\omega)$ for $\omega<2$
is equal to the initial energy $E_{0}(\omega)$ with the initial state
$|\theta_{0}=\frac{\pi}{2},\phi_{0}=0\rangle$, marked by the black
dashed line in Fig. (\ref{fig:E})(a). For a quantum system, $E_{c}(\omega)=\chi+\frac{N}{2}\omega$
and $E_{c}(\omega)/N=\frac{1}{2}\omega$ as $N\rightarrow\infty$
for $\omega\in(0,2)$. Meanwhile, under the semi-classical approximation,
we can obtain $E_{c}(\omega)=\frac{N}{2}\omega$, consistent with
the result of the quantum system in the thermodynamic limit.

\begin{figure}
\begin{centering}
\includegraphics[scale=0.17]{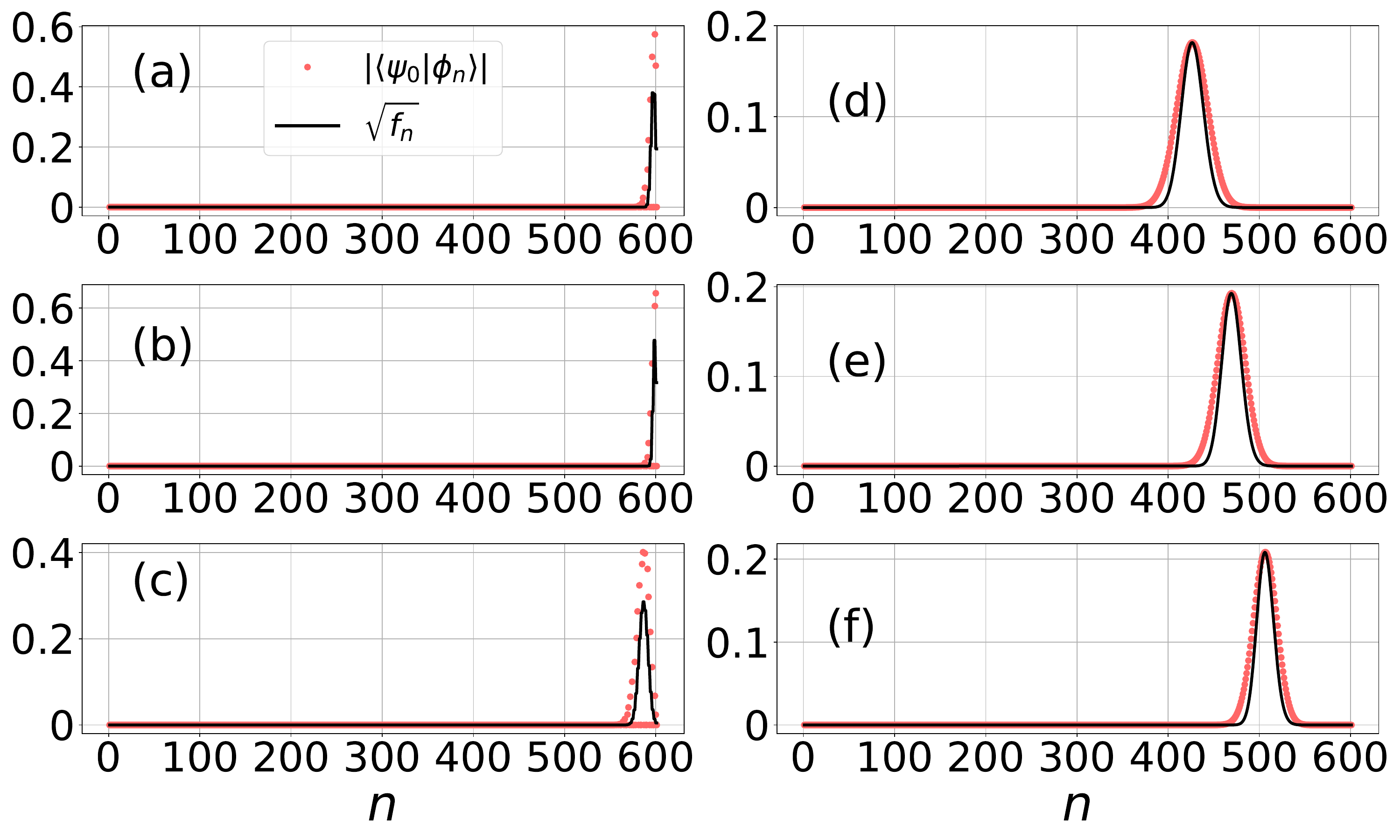}
\par\end{centering}
\caption{Distribution of $\left|\langle\psi_{0}|\phi_{n}\rangle\right|$ for
$N=600$. The parameters are $\omega=0.5$ for (a)(b)(c) and $\omega=2$
for (d)(e)(f). The initial states are (a)(d) $|\theta_{0}=0.05\pi,\phi_{0}=0.05\pi\rangle$;
(b)(e) $|\theta_{0}=0.1\pi,\phi_{0}=0.05\pi\rangle$; (c)(f) $|\theta_{0}=0.15\pi,\phi_{0}=0.05\pi\rangle$.\protect\label{fig:alpha_0j}}
\end{figure}

\begin{figure}
\begin{centering}
\includegraphics[scale=0.17]{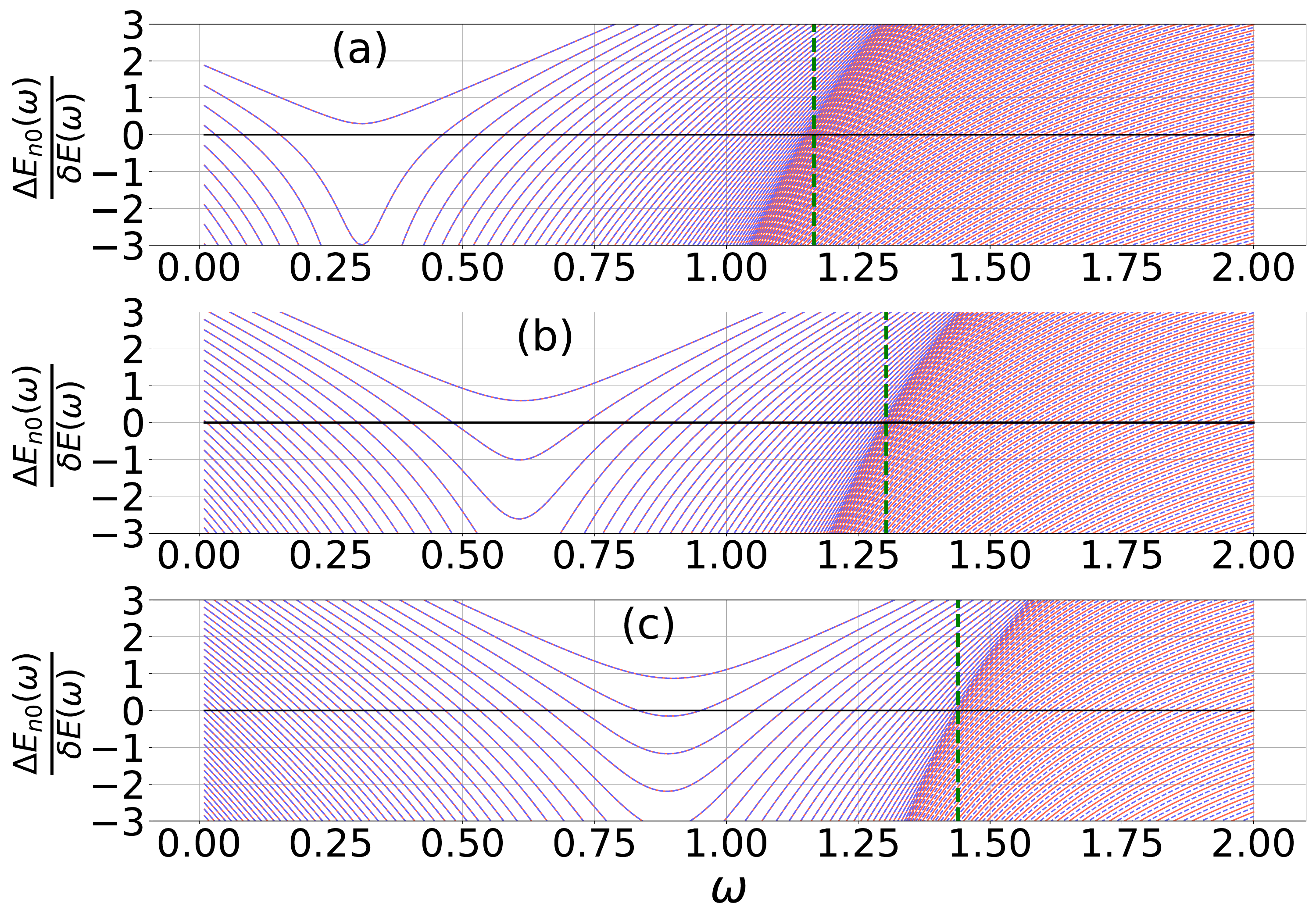}
\par\end{centering}
\caption{$\Delta E_{n0}/\delta E$ vs $\omega$ for $N=600$. The initial
states are (a) $|\theta_{0}=0.05\pi,\phi_{0}=0.05\pi\rangle$; (b) $|\theta_{0}=0.1\pi,\phi_{0}=0.05\pi\rangle$;
(c) $|\theta_{0}=0.15\pi,\phi_{0}=0.05\pi\rangle$. The horizontal
solid lines guide the value of $\Delta E_{n0}=0$. The vertical dashed
lines guide the value of $\omega_{c}$ obtained from the order parameter
$\overline{z}$. \protect\label{fig:DEn0}}
\end{figure}

Now we introduce the energy uncertainty of the initial state $|\psi_{0}\rangle$,
which is calculated by
\begin{equation}
\left(\delta E(\omega)\right)^{2}=\langle\psi_{0}|H^{2}(\omega)|\psi_{0}\rangle-\left(\langle\psi_{0}|H(\omega)|\psi_{0}\rangle\right)^{2}.
\end{equation}
Then we construct the Gaussian function from $E_{0}$ and $\delta E$:
\begin{equation}
f_{n}=\frac{1}{\mathcal{N}}e^{-\frac{\left[E_{n}(\omega)-E_{0}(\omega)\right]^{2}}{2\left[\delta E(\omega)\right]^{2}}},
\end{equation}
where $\mathcal{N}$ is the normalized coefficient. The quantity $f_{n}$
gives the information of how the initial state distributes within
eigenstates of the underlying Hamiltonian $H(\omega)$. We plot the
function $\sqrt{f_{n}}$ and the coefficients $\left|\alpha_{0n}\right|=\left|\langle\psi_{0}|\phi_{n}\rangle\right|$
versus $n$ in Fig. \ref{fig:alpha_0j}. It can be found that $\sqrt{f_{n}}$
almost recovers the distribution of $\left|\alpha_{0n}\right|$, indicating
that the distribution of $\left|\alpha_{0n}\right|$ is similar to
the Gaussian function with the center located at $E_{0}(\omega)$.
Also, we calculate the Eq. (\ref{eq:=000020Ckb_ana}) within the energy
window $\epsilon\in[E_{0}-2\delta E,E_{0}+2\delta E]$ and present
the results in Fig. \ref{fig:traj_LMG}(d) by dashed lines. The results
fit very well with the original data and capture the behaviour of
the transition. The normal distribution structure of the probability
density function $\left|\alpha_{0j}\right|^{2}$ means the behaviour
of $\overline{C}_{K}$ is dominated by a small portion of eigenstates
with eigenvalues near the initial energy. Focusing on the part of the
spectrum near the initial energy $E_{0}(\omega)$, we consider the
shifted spectrum $\Delta E_{n0}(\omega)=E_{n}(\omega)-E_{0}(\omega)$
with the unit of $\delta E(\omega)$ and display it in Fig. \ref{fig:DEn0}.
It can be observed that the structure of the energy spectrum changes from
two-fold degenerate region to non-degenerate region within the energy
window as $\omega$ increases. The transition point $\omega_{c}$
indicated by the dashed line is around the cross point of $E_{0}(\omega)$
and $E_{c}(\omega)$. Since $E_{0}(\omega)$ is dependent on the initial
state, its cross point with $E_{c}(\omega)$ depends on the initial
state too (see Fig. \ref{fig:DEn0}(a)). This gives an explanation why the dynamical
phase transition point $\omega_{c}$ is initial-state-dependent from
the perspective of spectrum structure.

Similar to the case of the LMG model, a symmetry-breaking transition of
eigenstates in the two-mode BECs can be triggered by the excited state
quantum phase transitions. For the doubly degenerate eigenstate $|\phi_{n}\rangle$,
we can adopt the notion of the partial symmetry introduced in the
study of the excited state quantum phase transition \citep{Corps2022PRB,Corps2023PRL},
with the partial symmetry operator defined as $\hat{\Pi}=\text{sign}(S_{z})$.
The partial symmetry operator is a $\mathbb{Z}_{2}$ operator, which
fulfills $\hat{\Pi}|\phi_{n}\rangle=\pm|\phi_{n}\rangle$.

\begin{figure}[H]
\begin{centering}
\includegraphics[scale=0.17]{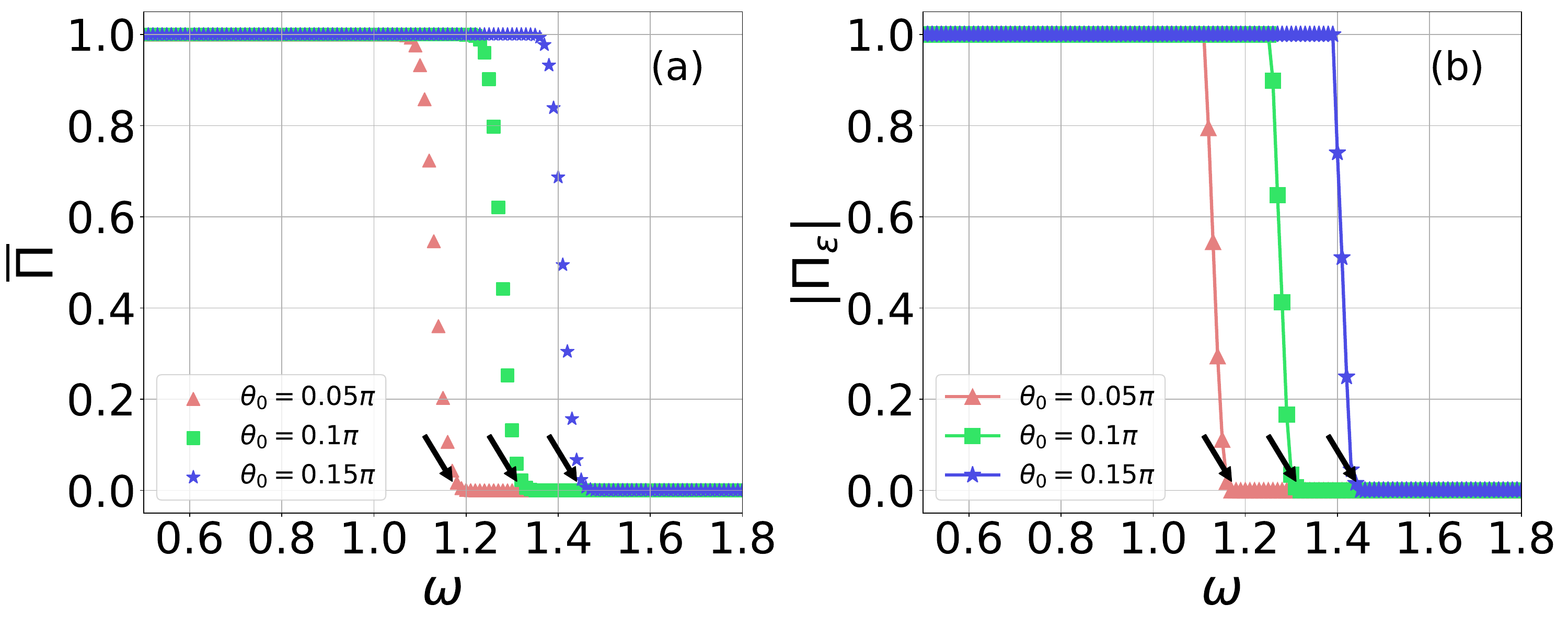}
\par\end{centering}
\caption{(a) $\overline{\Pi}$ and (b) $\overline{\Pi}_{\epsilon}$ vs
$\omega$ with $\phi_{0}=0.05\pi$ for $N=2000$. The arrows denote
the transition points obtained from the dynamical order parameter
$\overline{z}$. \protect\label{fig:SymO}}
\end{figure}

The time evolved state can be expanded in the eigenstates of the Hamiltonian:
%\textcolor{red}{}
\begin{equation}
|\psi(t)\rangle=\sum_{n=1}^{\mathcal{D}}e^{-iE_{n}t}\alpha_{0n}^{*}|\phi_{n}\rangle.
\end{equation}
Since our initial state satisfies $\langle\psi_{0}|\hat{\Pi}|\psi_{0}\rangle=1$,
when $\omega<\omega_{c}$, the time evolved state is restricted in
the one of two symmetry subspaces, and thus $\langle\hat{\Pi}\rangle$
is conserved. On the contrary, as the parameter $\omega$ crosses the
transition point, $\langle\hat{\Pi}\rangle$ is not conserved. To
see it clearly, we numerically calculate the long-time average of
the operator $\hat{\Pi}$:
\[
\overline{\Pi}=\sum_{n=1}^{\mathcal{D}}\left|\alpha_{0n}\right|^{2}\langle\phi_{n}|\hat{\Pi}|\phi_{n}\rangle,
\]
and the average value of $\left|\langle\phi_{n}|\hat{\Pi}|\phi_{n}\rangle\right|$
within the energy window $\epsilon\in[E_{0}-\delta E,E_{0}+\delta E]$,
which can be expressed as
\begin{equation}
\overline{\Pi}_{\epsilon}=\frac{1}{N_{\epsilon}}\sum_{E_{n}\in\epsilon}\left|\langle\phi_{n}|\hat{\Pi}|\phi_{n}\rangle\right|
\end{equation}
where $N_{\epsilon}$ is the number of eigenstates in the energy window
$\epsilon$. The values of $\overline{\Pi}$ and $\overline{\Pi}_{\epsilon}$
with respect to $\omega$ are shown in Fig. \ref{fig:SymO}. The transition
behaviour presented in $\overline{\Pi}$ and $\overline{\Pi}_{\epsilon}$
is consistent with the results of Fig.1(c) and Fig.1(d). For $\omega<\omega_{c}$,
both $\overline{\Pi}$ and $\overline{\Pi}_{\epsilon}$ equal to 1
as $\left|\alpha_{0n}\right|$ populates within the broken symmetry
state. On the other hand, they approach zero for $\omega>\omega_{c}$.

\section{Characterizing the transition between Chaotic and regular dynamics in triple-well bosonic systems}

\begin{figure}[H]
\begin{centering}
\includegraphics[scale=0.15]{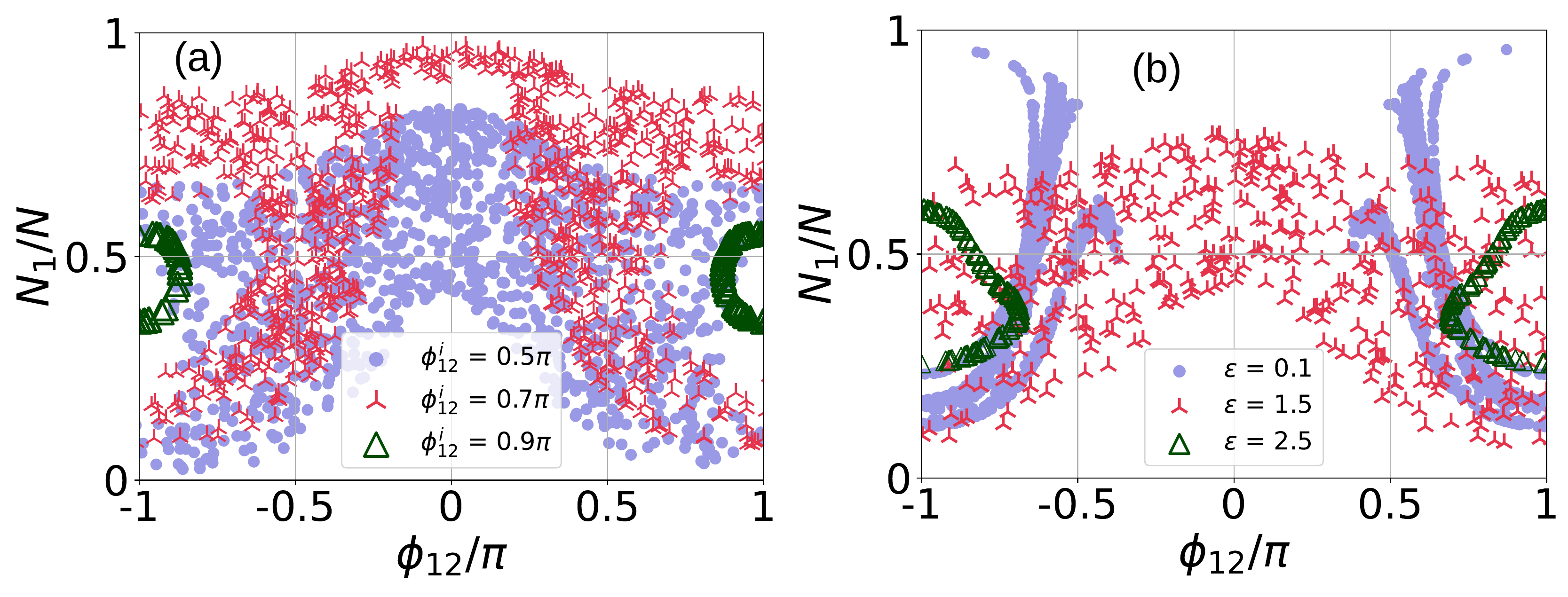}
\par\end{centering}
\caption{The projected Poincar\'{e} section spanned in the coordinate space of
$(N_{1},\phi_{12})$ with $\phi_{23}=0$. (a)  The parameter $\varepsilon=0.7$
and the initial state is $|N_{1}^{i}/N=0.4,N_{3}^{i}/N=0.3,\phi_{23}^{i}=0\rangle$. Three different symbols correspond to $\phi_{12}^{i}=0.5\pi,0.7\pi$, and $0.9\pi$.(b) The initial state is $|N_{1}^{i}/N=0.4,N_{3}^{i}/N=0.3,\phi_{12}^{i}=0.7\pi,\phi^i_{23}=0\rangle$. Three different symbols correspond to $\varepsilon=0.1,1.5$, and $2.5$.  In both cases, the parameters are $U=0.7$ and
$J=1$. \protect\label{fig:traj_triBEC}}
\end{figure}

To show that the $\overline{C}_{K}$ can distinguish more complex quantum dynamics, especially when the system displays chaotic dynamics and regular dynamics,  next we consider the triple-well bosonic model as an example system,
which is described by the Hamiltonian \citep{Rautenberg2020PRA,Santos2022PRE,Castro2024PRA}:
\begin{align}
H= & \frac{U}{N}\left(\hat{N}_{1}-\hat{N}_{2}+\hat{N}_{3}\right)^{2}+\varepsilon\left(\hat{N}_{3}-\hat{N}_{1}\right)\nonumber \\
 & +\frac{J}{\sqrt{2}}\left(\hat{a}_{1}^{\dagger}\hat{a}_{2}+\hat{a}_{2}^{\dagger}\hat{a}_{1}+\hat{a}_{2}^{\dagger}\hat{a}_{3}+\hat{a}_{3}^{\dagger}\hat{a}_{2}\right),
\end{align}
where $\hat{a}_{k}^{\dagger}$($\hat{a}_{k}$) is the creation (annihilation)
operator of the $k$th well and $\hat{N}_{k}=\hat{a}_{k}^{\dagger}\hat{a}_{k}$.
The total particle number $N=N_{1}+N_{2}+N_{3}$ is fixed. For large enough $N$, we can replace the operators $\hat{a}_{k}\rightarrow \sqrt{N_{k}}e^{i \phi_k}$, then the classical Hamiltonian can be written as\citep{Castro2024PRA}
\begin{align}
\bar{H}_{cl}\equiv\frac{H_{\text{cl}}}{N}= & U(\rho_{1}^{2}-\rho_{2}^{2}+\rho_{3}^{2})^{2}+\varepsilon\left(\rho_{3}^{2}-\rho_{1}^{2}\right)\nonumber \\
 & +J\sqrt{2}\left[\rho_{1}\rho_{2}\cos\phi_{12}+\rho_{2}\rho_{3}\cos\phi_{23}\right], \label{eq:Hcl_triBEC}
\end{align}
where $\rho_{k}\equiv\sqrt{N_{k}/N}$ and $\phi_{jk}\equiv\phi_{j}-\phi_{k}$.
$U$ is the magnitude of the boson interaction, $J$ parameterizes
the jump between wells, and $\varepsilon$ is an external tilt. Introducing the canonical variables:
\begin{equation}
\begin{cases}
Q_{1}=\sqrt{2}\rho_{1}\cos\phi_{1},P_{1}=\sqrt{2}\rho_{1}\sin\phi_{1},\\
Q_{2}=\sqrt{2}\rho_{2}\cos\phi_{2},P_{2}=\sqrt{2}\rho_{2}\sin\phi_{2},\\
Q_{3}=\sqrt{2}\rho_{3}\cos\phi_{3},P_{3}=\sqrt{2}\rho_{3}\sin\phi_{3},
\end{cases}
\end{equation}
 the classical Hamiltonian can be written as follows:
\begin{align}
\bar{H}_{cl}=	&\frac{U}{4}(Q_{1}^{2}+P_{1}^{2}-Q_{2}^{2}-P_{2}^{2}+Q_{3}^{2}+P_{3}^{2})^{2}  \nonumber \\
	&+\frac{\epsilon}{2}(Q_{3}^{2}+P_{3}^{2}-Q_{1}^{2}-P_{1}^{2})  \nonumber \\
	&+\frac{J}{\sqrt{2}}\left(Q_{1}Q_{2}+P_{1}P_{2}+Q_{2}Q_{3}+P_{2}P_{3}\right).  \label{eq:Hcl_triBEC_ca1}
\end{align}
The equation of motion are given by
\begin{equation}
\begin{cases}
\dot{Q}_{i}=\frac{\partial\bar{H}_{cl}}{\partial P_{i}},\\
\dot{P}_{i}=-\frac{\partial\bar{H}_{cl}}{\partial Q_{i}},  \label{eq:eom_triBEC_ca1}
\end{cases}
\end{equation}
with $i\in\{1,2,3\}$. As the total number of bosons $N=N_1+N_2+N_3$, we can substitute
$\rho_2=\sqrt{1-\rho_1^2-\rho_3^2}$ into Eq. (\ref{eq:Hcl_triBEC}). Then the classical Hamiltonian can be brought into a simpler form:
\begin{align}
\bar{H}_{cl}=	&U(q_{1}^{2}+p_{1}^{2}+q_{3}^{2}+p_{3}^{2}-1)^{2} \nonumber \\
	&+\frac{\epsilon}{2}(q_{3}^{2}+p_{3}^{2}-q_{1}^{2}-p_{1}^{2}) \nonumber \\
	&+J(q_{1}+q_{3})\sqrt{1-\frac{q_{1}^{2}+p_{1}^{2}+q_{3}^{2}+p_{3}^{2}}{2}},   \label{eq:Hcl_triBEC_ca2}
\end{align}
where 
\begin{equation}
\begin{cases}
q_{1}=\sqrt{2}\rho_{1}\cos\phi_{12},p_{1}=\sqrt{2}\rho_{1}\sin\phi_{12},\\
q_{3}=\sqrt{2}\rho_{3}\cos\phi_{23},p_{3}=\sqrt{2}\rho_{3}\sin\phi_{23},
\end{cases}
\end{equation}
are the new canonical variables. The classical Hamiltonians Eq. (\ref{eq:Hcl_triBEC_ca1}) and Eq. (\ref{eq:Hcl_triBEC_ca2}) are equivalent. However, $\dot{Q}_i$ and $\dot{P}_i$ are simple polynomials in the three pairs of $(Q_k, P_k)$ variables. For numerical calculations, it is more efficient to use Eq. (\ref{eq:eom_triBEC_ca1}) to calculate the classical trajectories. In Fig. \ref{fig:traj_triBEC}, we focus on the projected Poincar\'{e} section  spanned in the coordinate space of $(N_{1},\phi_{12})$ with $\phi_{23}=0$.   The coordinate space of $(N_{1},\phi_{12})$ corresponds to the pair of canonically conjugate variables $(q_1,p_1)$.
This model exhibits different dynamical behaviour depending on the
choice of the parameters and the initial state. In the case of $U=0.7,\varepsilon=0.7$ and $J=1$, we have the coexistence of  regular and chaotic regions present in the phase space\citep{Castro2024PRA}. We illustrate the classical trajectories for three different initial states
in Fig. \ref{fig:traj_triBEC}(a).  It can be seen that the trajectory displays chaotic
dynamics for $\phi_{12}^{i}=0.5\pi$ and regular dynamics for $\phi_{12}^{i}=0.9\pi$.
Additionally, we show the classical trajectories for different parameter $\varepsilon$
with fixed initial states in Fig. \ref{fig:traj_triBEC}(b), where the
change from chaotic $\left(\varepsilon=1.5\right)$ to regular
dynamics $\left(\varepsilon=2.5\right)$ can be observed.  It should be noted that the trajectory is regular-like when the system slightly deviates from the integrable point  ($\varepsilon=0$), such as $\varepsilon=0.1$ in Fig. \ref{fig:traj_triBEC}(b).

\begin{figure}[H]
\begin{centering}
\includegraphics[scale=0.15]{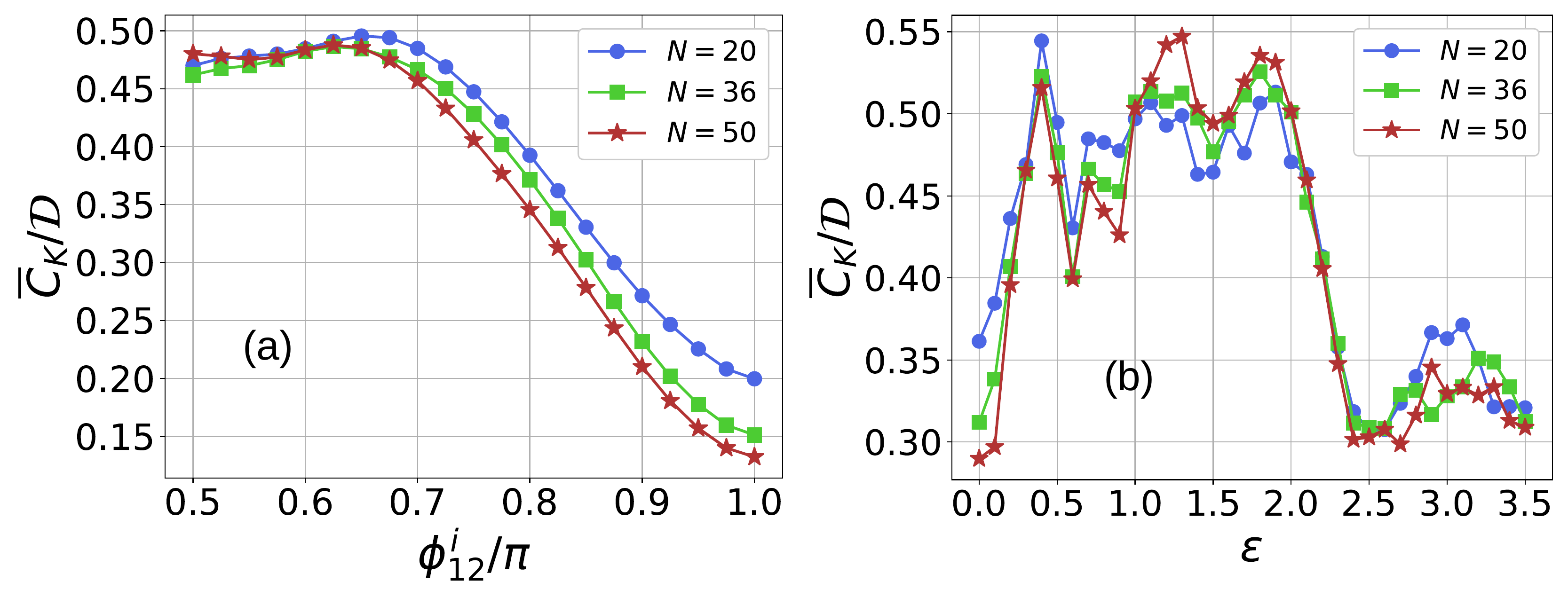}
\par\end{centering}
\caption{ $\overline{C}_{K}/\mathcal{D}$ vs  $\phi_{12}^{i}$ and
$\varepsilon$ for different total particle number, where $\mathcal{D}=\frac{(N+2)(N+1)}{2}$
is the dimension of the Hilbert space. (a) The parameter $\varepsilon=0.7$
and the initial state is $|N_{1}^{i}/N=0.4,N_{3}^{i}/N=0.3,\phi_{23}^{i}=0\rangle$. (b) The initial state is $|N_{1}^{i}/N=0.4,N_{3}^{i}/N=0.3,\phi_{12}^{i}=0.7\pi, \phi^i_{23}=0\rangle$. In both cases, the parameters are $U=0.7$ and
$J=1$. \protect\label{fig:QC_triBEC}}
\end{figure}

For quantum dynamics, we choose the coherent states as the initial state\citep{Castro2024PRA}:
\begin{align}
|\psi_{0}\rangle= & |N_{1}^{i},N_{3}^{i},\phi_{12}^{i},\phi_{23}^{i}\rangle\nonumber \\
= & \sum_{n_{1}+n_{2}+n_{3}=N}\sqrt{P}e^{in_{1}\phi_{12}^{i}}e^{in_{3}\phi_{23}^{i}}|n_{1},n_{2},n_{3}\rangle,
\end{align}
where $P=\frac{N!}{n_{1}!n_{2}!n_{3}!}p_{1}^{n_{1}}p_{2}^{n_{2}}p_{3}^{n_{3}}$
is the multinomial distribution with $p_{k}=\frac{N_{k}^{i}}{N}$. In Fig. \ref{fig:QC_triBEC}(a),
we calculate the long-time average of the spread complexity $\overline{C}_{K}$
with the initial state varying from $\phi_{12}^{i}=0.5\pi$ to $\pi$
and other conditions the same as in Fig. \ref{fig:traj_triBEC}(a). The $\overline{C}_{K}$ displays a transition
from higher values to lower values corresponding to the dynamical
transition from chaotic to regular dynamics. In addition, we vary the parameter
$\varepsilon$ from zero to $3.5$ with other conditions the same as Fig. \ref{fig:traj_triBEC}(b) and
show the value of $\overline{C}_{K}$ in Fig. \ref{fig:QC_triBEC}(b).
The system is integrable at $\varepsilon=0$ and the regular dynamics
appear in the phase space. Therefore, the value of $\overline{C}_{K}$
increase from a small value when $\varepsilon$ increase from 0. As
the system enter the regime of high degree of chaos $\left(\varepsilon\in[1.2,1.7]\right)$\citep{Santos2022PRE},
the value of $\overline{C}_{K}$ approaches to a higher value, indicating
that the quantum state highly spreads out in the phase
space. Further increasing the parameter $\varepsilon$,  regular
dynamics reappears in the phase space which can be observed in Fig. \ref{fig:traj_triBEC}(d) with
$\varepsilon=2.5$.  Consequently, the value of $\overline{C}_{K}$ drops to
a small value which is close to the integrable case.

\section{Summary}

In summary, we have studied the spread complexity $C_{K}$ and its
long-time average value $\overline{C}_{K}$ in the two-mode BECs.
Our results demonstrate that the long-time average of the spread complexity
$\overline{C}_{K}$ can probe the dynamical transition in the two-mode
BECs. By choosing the spin coherent state as the initial state, we find
that $\overline{C}_{K}$ exhibits a sharp transition as the phase
space trajectory of the time evolved state changes from the self-trapping
to Josephson oscillation. By examining the eigen-spectrum of the underlying
Hamiltonian, we identified the existence of an excited state quantum
phase transition in the region of $\omega<2$, characterized by the
emergence of singularity in the density of states at critical energy
$E_{c}$. In the thermodynamical limit, the critical energy separates
doubly degenerate eigenstates from non-degenerate eigenstates. We
unraveled that the dynamical transition point is determined by the
cross point of the initial energy $E_{0}(\omega)$ and $E_{c}(\omega)$.
When $\omega$ exceeds a threshold $2$, the fixed point $\left(\theta=\frac{\pi}{2},\phi=0\right)$
changes from a saddle point to a stable fixed point. By studying the
dynamics for the initial state at this fixed point, we unveiled that
the different dynamical behavior in the region of $\omega<2$ and
$\omega>2$ can be distinguished by the long-time average of the spread
complexity. Also, we have studied $\overline{C}_{K}$
in the triple-well bosonic model. Our results imply that the long-time
average of the spread complexity can detect the transition
from chaotic dynamics to regular dynamics.
\begin{acknowledgments}
This work is supported by National Key Research and Development Program
of China (Grant No. 2021YFA1402104 and 2023YFA1406704), the NSFC under
Grants No. 12174436 and No. T2121001, and the Strategic Priority Research
Program of Chinese Academy of Sciences under Grant No. XDB33000000.
\end{acknowledgments}

\begin{widetext}
\appendix

\section{Husimi function of the Krylov basis}

In the main text, we show that the spread complexity is directly connected
to the area in which the trajectory spreads in the phase space. Here,
we give more evidence to this connection by demonstrating the distribution of Husimi function
$\mathcal{H}_{k}(\theta,\phi)$ in the phase space for the $k$th
Krylov basis $|K_{k}\rangle$. The $\mathcal{H}_{k}(\theta,\phi)$
is defined as
\begin{equation}
\mathcal{H}_{k}(\theta,\phi)=|\langle\theta,\phi|K_{k}\rangle|^{2},
\end{equation}
where $|\theta,\phi\rangle=e^{-iS_{z}\phi}e^{-iS_{y}\theta}|\frac{N}{2},\frac{N}{2}\rangle$
is the coherent spin state. To gain an intuitive insight, we illustrate distributions of $\mathcal{H}_{k}(\theta,\phi)$
for several Krylov bases in Fig. \ref{fig:Kry_coh}. It can be clearly seen that different
Krylov bases are associated with different regions in the phase space. It can also be found that the Krylov basis gradually spreads out  from the location of the initial state as $k$ increases from zero. Thus, a larger value of $\overline{C}_K$ corresponds to a larger area over which the quantum state evolves.
\begin{figure}[H]
\begin{centering}
\includegraphics[scale=0.42]{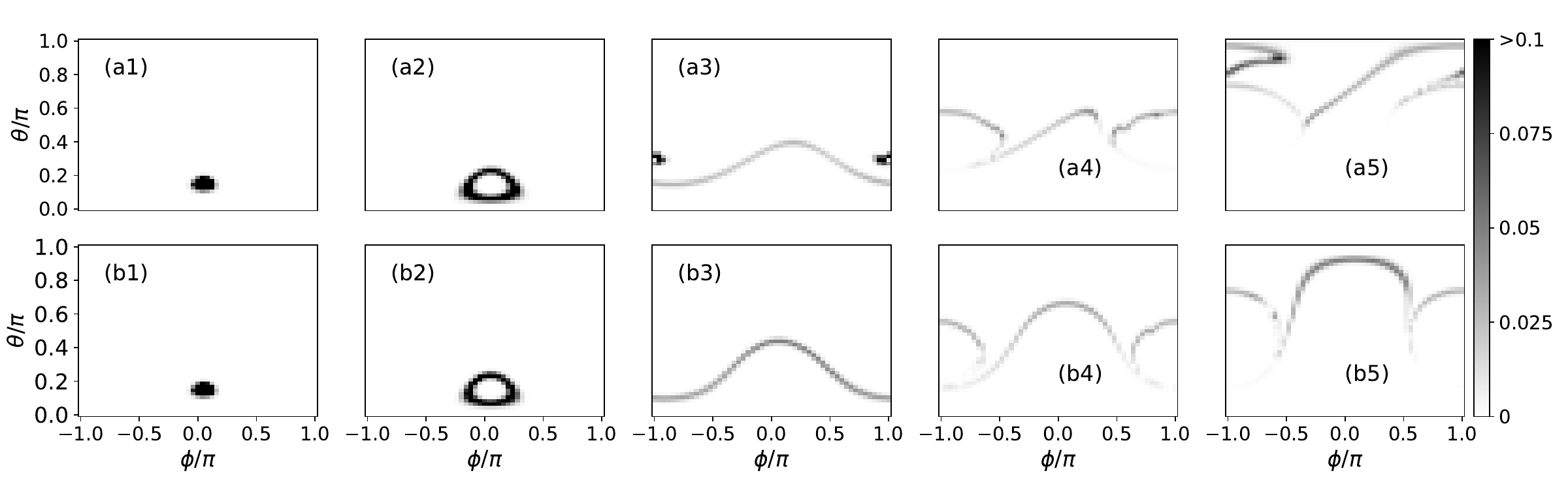}
\par\end{centering}
\caption{$\mathcal{H}_{k}(\theta,\phi)$ for (a1)(b1) $k=0$; (a2)(b2) $k=10$;
(a3)(b3) $k=100$; (a4)(b4) $k=300$; (a5)(b5) $k=500$. The parameters are (a1)$\sim$(a5) $\omega=0.5$ and (b1)$\sim$(b5) $\omega=2$.
 The initial state is $|\theta_{0}=0.15\pi,\phi_{0}=0.05\pi\rangle$ with particle number $N=600$.
\protect\label{fig:Kry_coh}}
\end{figure}

\section{Inverse Participation ratio and return probability of the time evolved state in the Krylov
space }

To show the wave function spreading out, we calculate the inverse
participation ratio of the time evolved state given by
$\mathcal{I}_{K}(t)=\sum_{k}\left|\langle K_{k}|\psi(t)\rangle\right|^{4}$.
It can be seen from Figs. \ref{fig:IPRt_ome}(a)(b) that the $\mathcal{I}_{K}(t)$
saturates to a very small value at late times for $\omega=2$ which
corresponds to a higher value of $\overline{C}_{K}$. A small
value of $\mathcal{I}_{K}(t)$ indicates that the state is
delocalized. In contrast, $\mathcal{I}_{K}(t)$ saturates to
a larger value for $\omega=0.5$, with the corresponding  $\overline{C}_{K}$ displaying a lower value.
In this case, we note that the value of $\mathcal{I}_{K}(t)$ periodically approaches to 1, which results from the
state periodically approaching the initial state. This can be observed
in the return probability $\mathcal{L}(t)=|\langle\psi_{0}|\psi(t)\rangle|^{2}$
(Figs. \ref{fig:IPRt_ome}(c)(d)).

\begin{figure}[H]
\begin{centering}
\includegraphics[scale=0.28]{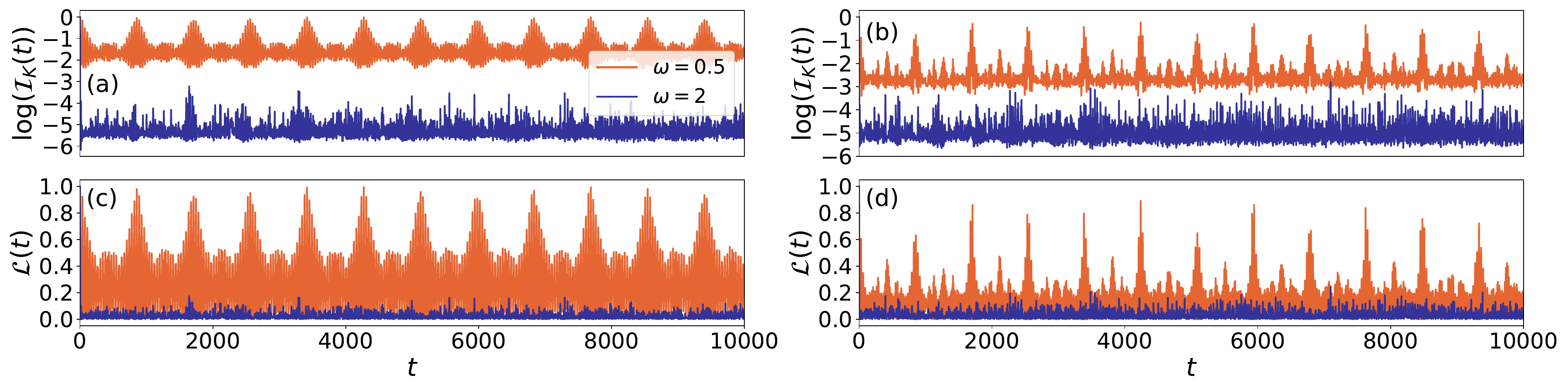}
\par\end{centering}
\caption{(a)(b) $\mathcal{I}_{K}(t)$ and (c)(d) $\mathcal{L}(t)$ for
$\omega=0.5$ and $2$. The initial states are (a)(c) $|\theta_{0}=0.05\pi$ $,\phi_{0}=0.05\pi\rangle$;
(b)(d) $|\theta_{0}=0.15\pi$ $,\phi_{0}=0.05\pi\rangle$.\protect\label{fig:IPRt_ome} }
\end{figure}
\end{widetext}

\end{document}